%% file: LesHouchesSethna06.tex
\documentclass{article}
\usepackage{bookstyle,bm,cite}
\usepackage{amsmath,amssymb,amsfonts,epsfig,url,citesort}
\begingroup
\makeatletter
\g@addto@macro{\UrlSpecials}{%
  \endlinechar=13 \catcode\endlinechar=12
  \do\%{\Url@percent}\do\^^M{\break}}
 \gdef\Url@percent{\@ifnextchar^^M{\@gobble}{\mathbin{\mathchar`\%}}}%
\endgroup %

\input{JPSCommands}

\begin{document}
\title{Crackling Noise and Avalanches: Scaling, Critical Phenomena, and the
Renormalization Group}
\runtitle{Crackling Noise and Avalanches}
\author{James P. Sethna}
\address{Laboratory of Atomic and Solid State Physics, Cornell University,\\ Ithaca, NY, USA}
\frontmatter
\maketitle
\mainmatter%

\section{Preamble}

In the past two decades or so, we have learned how to understand crackling
noise in a wide variety of systems. We review here the basic ideas and
methods we use to understand crackling noise---critical phenomena,
universality, the renormalization group, power laws, and universal scaling
functions. These methods and tools were originally developed to
understand continuous phase transitions in thermal and disordered systems, 
and we also introduce these more traditional applications as illustrations
of the basic ideas and phenomena.

We focus largely on crackling noise in magnetic hysteresis, called
{\em Barkhausen noise}. These lecture notes are distilled from a review article
written with Karin Dahmen and Christopher Myers~\cite{SethDahmMyer01},
from a book chapter written with Karin Dahmen and 
Olga Perkovi{\,c}~\cite{SethDahmPerk06}, and from a chapter in my
textbook~\cite{Seth06}.

\section{What is crackling noise?}
\label{sec:WhatIsCracklingNoise}

\begin{figure}[bth]
  \begin{center}
    \epsfxsize=\hsize
    \epsffile{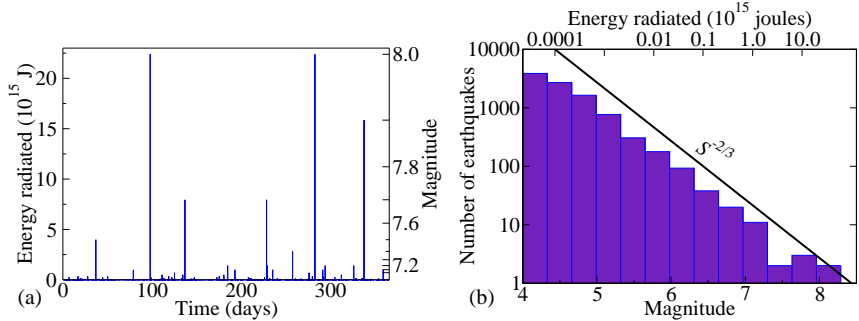}
  \end{center}
\caption{{\bf Earthquake sizes.} 
  (a)~Earthquake energy release versus time in 1995. There are only a few
  large earthquakes, and many small ones.
  This time series, when sped up, sounds like
  crackling noise~\cite{url:CracklingNoise}.
  (b)~Histogram of the number of earthquakes in 1995 as a function of
  their magnitude $M$. Notice the
  logarithmic scales; the smallest earthquakes shown are a million times 
  smaller and a thousand times more probable than the largest earthquakes. The
  fact that the earthquake size distribution is well described by a power 
  is the Gutenberg--Richter law~\cite{GutenRich}.
}
\label{fig:Earthquakes}
\end{figure}

\begin{figure}[thb]
  \begin{center}
    \null\hskip -0.3truein
    \epsfxsize=\hsize
    \epsffile{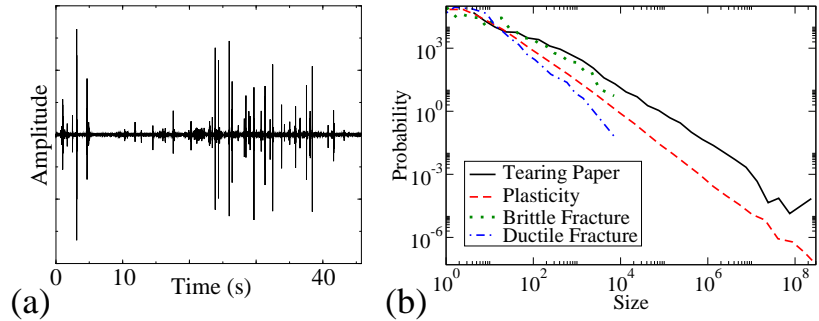}
  \end{center}
\caption{(a)~{\bf Crackling noise in paper} $V(t)$ for multiple pulses,
from~\cite{HoulSeth96}. Again, there are many small pulses, and only a 
few large events.
(b)~{\bf Power laws} of event sizes for paper 
tearing~\cite{SalmTolvAlav02}, plastic flow in ice~\cite{MiguVespZapp01},
and brittle and ductile fracture in steel~\cite{KunLenkTaka04}.
}
\label{fig:CracklingSizes}
\end{figure}

\begin{figure}[thb]
  \begin{center}
    \epsfxsize=\hsize
    \epsffile{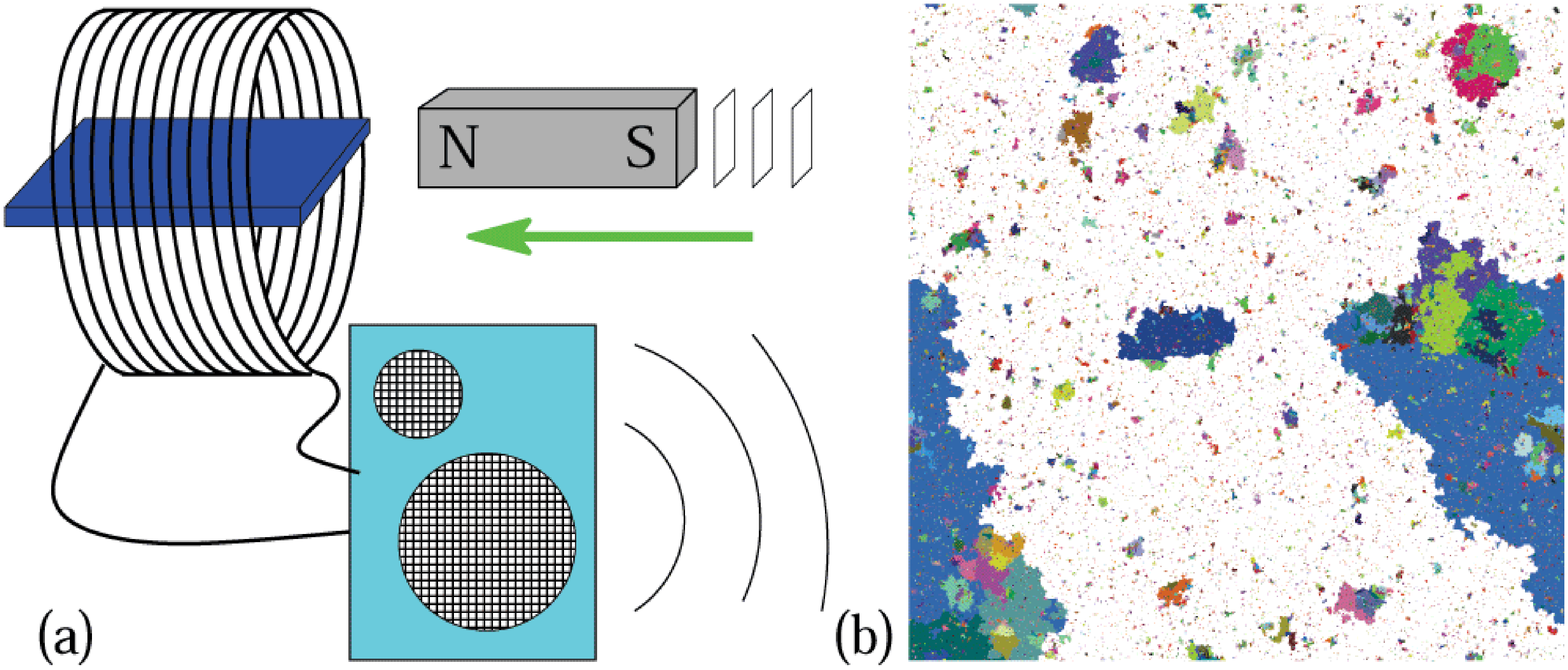}
  \end{center}
\caption{(a)~{\bf Barkhausen noise experiment.} By increasing an external
magnetic field $H(t)$ (bar magnet approaching), the magnetic domains in 
a slab of
iron flip over to align with the external field. The resulting magnetic field
jumps can be turned into an electrical signal with an inductive coil, and
then listened to with an ordinary loudspeaker. Barkhausen
noise from our computer experiments can be heard on the
Internet~\cite{url:CracklingNoise}.
(b)~{\bf Cross section of all avalanches} in a billion--domain simulation
of our model for Barkhausen noise at the critical
disorder~\cite{SethDahmMyer01}. The white background is the
infinite, spanning avalanche.
}
\label{fig:BarkhausenIntro}
\end{figure}

Many systems, when stressed or deformed slowly, respond with discrete events
spanning a broad range of sizes. We call this {\em crackling noise}. The
Earth crackles, as the tectonic plates rub past one another.
The plates move in discrete earthquakes (Fig.~\ref{fig:Earthquakes}a),
with many small earthquakes and only a few large ones. If the earthquake
series is played as an audio clip, sped up by a factor of ten million, it
sounds like crackling~\cite{url:CracklingNoise}. A piece of 
paper~\cite{HoulSeth96} or a candy wrapper~\cite{KramLobk96} will crackle
as it is crumpled (try it!), emitting sharp sound pulses as new creases
form or the crease pattern reconfigures (Fig.~\ref{fig:CracklingSizes}(a)).
Paper also tears in a series
of avalanches~\cite{SalmTolvAlav02}; fracture in many other systems 
exhibit avalanche distributions and jerky 
motion~\cite{KrysMayn98,MaloSantSchm+06}. Foams (a head of beer) move in 
jerky avalanches as they are sheared~\cite{TewaSchiDuri+99}, and as the
bubbles pop~\cite{VandLentDorb01}. Avalanches arise when fluids invade 
porous media in avalanches~\cite{CiepRobb88,NaraFish93} (such as water 
soaking into a sponge). I used to say that metals were an example of something
that did not crackle when bent (permanently plastically deformed), but 
there is now excellent data that ice crackles when it is 
deformed~\cite{MiguVespZapp01}, and recent data on micron-scale metal 
deformation also shows crackling noise~\cite{UchiDimiFlor04,DimiWoodLeSa06}.
We will focus here on {\em Barkhausen noise}, the magnetic pulses emitted
from (say) a piece of iron as it is placed inside an increasing external
field (see Fig.~\ref{fig:BarkhausenIntro}(a)).

All of these systems share certain common features. They all have many more
tiny events than large ones, typically
with a {\em power-law} probability distribution 
(Figs~\ref{fig:Earthquakes}(a),~\ref{fig:CracklingSizes}(a)). For example, 
the histogram of the number of earthquakes of a given size yields a straight
line on a log--log plot (Figure~\ref{fig:Earthquakes}b). This implies that
the probability of a large earthquake goes as its size (energy radiated) to
a power. The tearing avalanches in paper, dislocation avalanches in 
deformed ice, and rupture avalanches in steel all have power-law avalanche
size distributions(Fig.~\ref{fig:CracklingSizes}(b)). Because crackling noise
exhibits simple, emergent features (like these power law size distributions),
it encourages us to expect that a theoretical understanding might be 
possible.

What features of, say, an earthquake fault do we expect to be important
for such a theoretical model? If earthquake faults 
slipped only in large, snapping events we would expect to need to know
the shape of the tectonic plates in order to describe them. If they
slid more-or-less smoothly we anticipate that the nature of the internal
rubble and dirt (fault gouge) would be important. But since earthquakes
come in all sizes, we expect that neither the microscopic rock-scale nor the 
macroscopic continental-scale details can be crucial. What, then, is important
to get right in a model?

\begin{figure}[thb]
  \begin{center}
    \epsfxsize=\hsize
    \epsffile{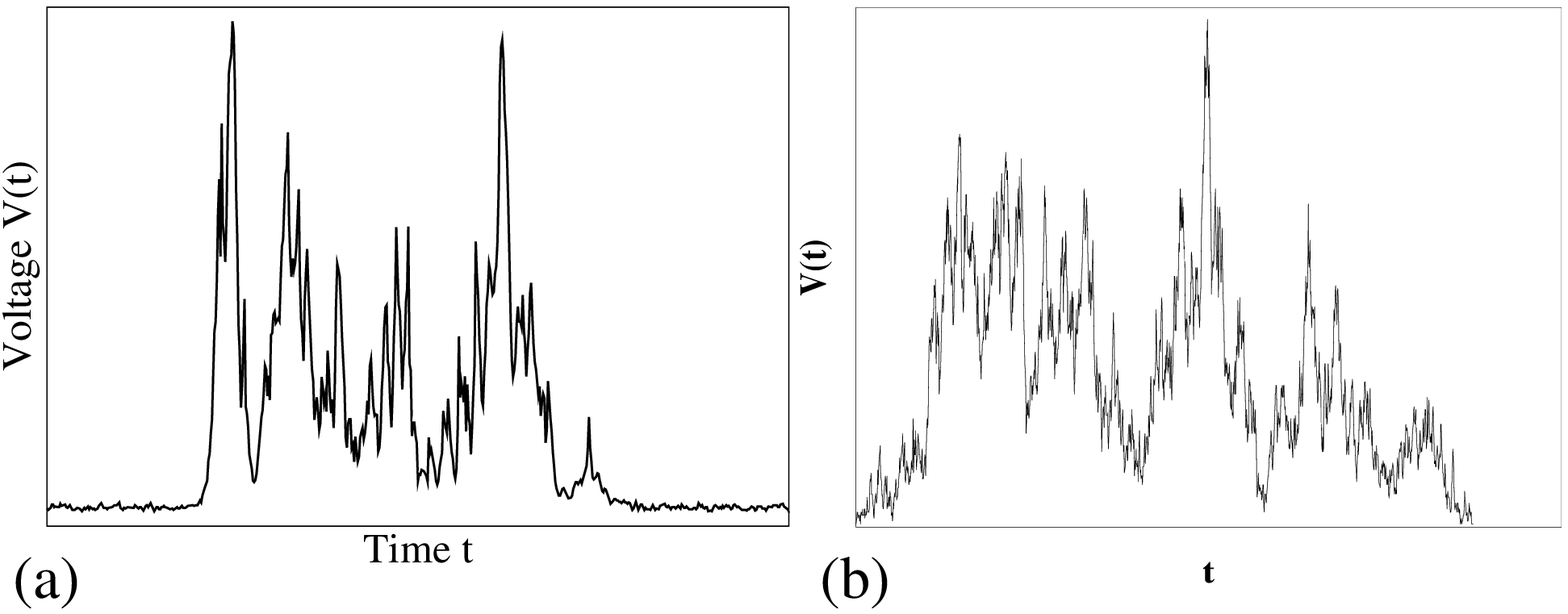}
  \end{center}
\caption{{\bf Internal avalanche time series} (a)~during a single
experimental Barkhausen noise pulse~\cite{ZappCastCola05}, and (b)~during 
a single avalanche in a simulation of magnetic Barkhausen
noise~\cite{SethDahmMyer01}. The experiment measures the voltage $V(t)$,
which in this experiment measures the volume swept out per unit time by
the moving magnetic domain wall; the simulation measures the number of
domains flipped per unit time. In these two cases, the total area under the
curve gives the size $S$ of the avalanche. Notice
how the avalanches almost stop several times; if the forcing were
slightly smaller, the large avalanche would have broken up into two or
three smaller ones. The fact that the forcing is just large enough to on
average keep the avalanche growing is the cause of the self-similarity;
a partial avalanche of size $S$ will on average trigger one other of size
$S$.}
\label{fig:TimeSeriesPulses}
\end{figure}

\begin{figure}[thb]
  \begin{center}
    \epsfxsize=\hsize
    \epsffile{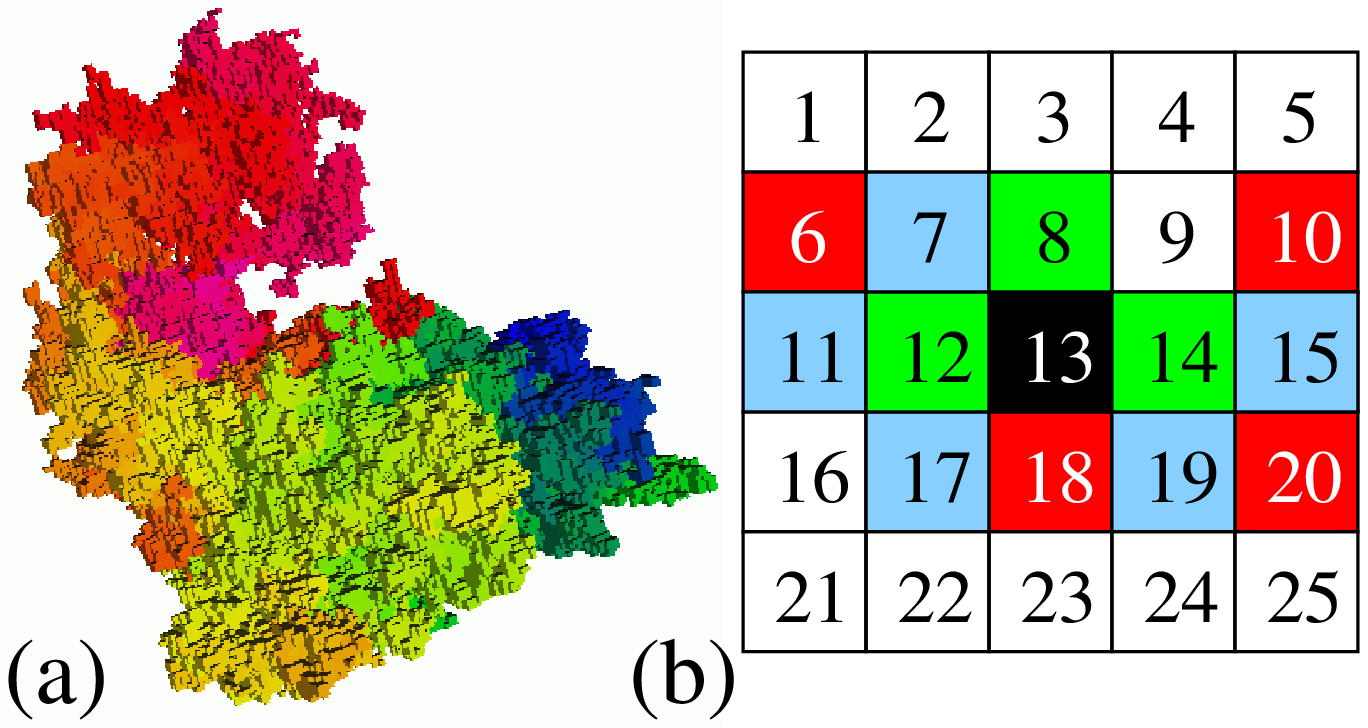}
  \end{center}
\caption{
(a)~{\bf Fractal spatial structure of an avalanche.} This moderate--sized
avalanche contains 282,785 domains (or spins)~\cite{SethDahmMyer01}.
The shading depicts the time evolution: the avalanche started in the
dark region in the back, and
the last domains to flip are in the upper, front region. The sharp changes
in shading are real, and represent sub-avalanches separated by times
where the avalanche almost stops (see Fig.~\ref{fig:TimeSeriesPulses}).
(b)~{\bf Avalanche propagation.} The avalanches in many models of
crackling noise are first nucleated when the external stress
induces a single site to transform.
In this two-dimensional model, site \#13 is
the nucleating site which is first pushed over. The coupling between
site \#13 and its neighbors triggers some of them to flip
(\#4, 8, and 12), which in turn trigger another shell of neighbors
(\#15, 19, 7, 11, and 17), ending eventually in a final shell which
happens not to trigger further sites (\#6, 10, 18, and 20). The time series
$V(t)$ plotted in figure~\ref{fig:TimeSeriesPulses}(b)
is just the number of domains flipped in shell \#t for that avalanche.
}
\label{fig:BigAndSmallAvalanches}
\end{figure}

An important hint is provided by looking at the dynamics of an individual
avalanche. In many (but not all) systems exhibiting crackling noise,
the avalanches themselves have complex, internal structures. 
Fig.~\ref{fig:TimeSeriesPulses} shows that the avalanches producing 
individual Barkhausen pulses in magnets proceed in an irregular, jerky
fashion, almost stopping several times in between sub-avalanches. 
Fig.~\ref{fig:BigAndSmallAvalanches}(a) shows that the spatial structure
of an avalanche in a model magnet is
also irregular. Both the time and spatial structures are {\em self-similar};
if we take one of the sub-avalanches and blow it up, it looks statistically
much like the original avalanche. Large avalanches are built up from multiple,
similar pieces, with each sub-avalanche triggering the next. It is the 
way in which one sub-avalanche triggers the next that
is crucial for a theoretical model to get right.

Our focus in these lecture notes will be on understanding the emergent behavior
in crackling noise (power laws and scaling functions) by exploring the
consequences of this self-similar structure. In section~\ref{sec:Barkhausen},
we illustrate the process by which models are designed and tested by
experiments using Barkhausen noise. In section~\ref{sec:WhyCrackling}
we introduce the {\em renormalization group}, and use it to explain
{\em universality} and self-similarity in these (and other) systems. 
In section~\ref{sec:PowerLawsScalingFunctions} we use the renormalization
group to explain the power laws characteristic of crackling noise, and 
also use it to derive the far more powerful {\em universal scaling functions}.

\section{Hysteresis and Barkhausen noise in magnets}
\label{sec:Barkhausen}

\begin{figure}[thb]
  \begin{center}
    \epsfxsize=\hsize
    \epsffile{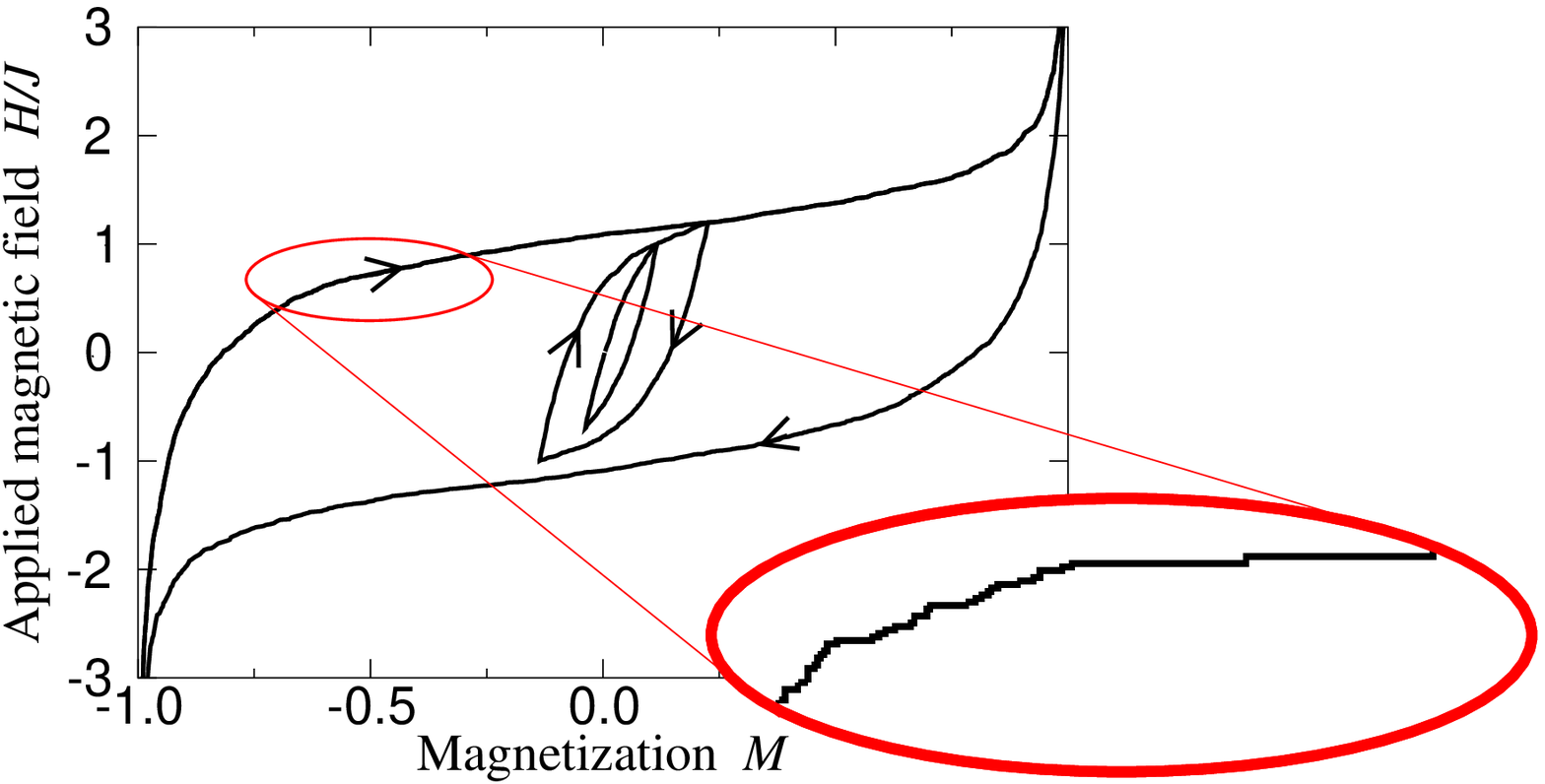}
  \end{center}
\caption{{\bf Hysteresis loop and subloops}: The magnetization in
our model~\cite{SethDahmKart93}, as the external field $H$ is ramped up
and down.
Our focus will primarily be on the upper, outer loop as the external field
is ramped from $-\infty$ to $\infty$. {\bf Barkhausen jumps} (exploded region):
The hysteresis loop appears smooth, but when examined in detail
is composed of discrete, abrupt jumps in magnetization, corresponding
to avalanches in the positions of the walls of the magnetic domains.
This jerky magnetization is what emits {\em Barkhausen noise}.
}
\label{fig:HysteresisLoopTinyJump}
\end{figure}

Microscopically, iron at room temperature is always magnetized; non-magnetic
bulk iron is composed of tiny magnetic domains, whose north poles point
in different directions giving a net zero magnetization. An external 
magnetic field, as in Fig.~\ref{fig:BarkhausenIntro}(a), will attract a 
piece of iron by temporarily moving the domain walls to enlarge the
regions whose north poles are aligned towards the south pole of the
external field. When the external field is removed, these domain walls
do not completely return to their original positions, and the iron will
end up partially magnetized; this history dependence is called 
{\em hysteresis}. The dependence of the magnetization of the 
iron $M$ on the history of the external field $H$ (the {\em hysteresis loop})
can have an interesting hierarchical structure of subloops 
(Fig.~\ref{fig:HysteresisLoopTinyJump}).

These hysteresis loops may look smooth, but when 
Fig.~\ref{fig:HysteresisLoopTinyJump} is examined in detail we see that
the magnetization grows in discrete jumps, or avalanches. These
avalanches are the origin of magnetic Barkhausen noise. In 
section~\ref{sbsec:RFIM} we will develop a simple model for this Barkhausen
noise. In section~\ref{sbsec:RealBarkhausen} we shall observe that 
our model, while capturing some of the right behavior, is not the correct
model for real magnets, and introduce briefly more realistic models
that do appear to capture rather precisely the correct behavior.%
  \footnote{That is, we believe they are in the right {\em universality
  class}, section~\ref{sbsec:Universality}.}

\subsection{Dynamical Random-field Ising model}
\label{sbsec:RFIM}

We introduce here a caricature of a magnetic material. We model the iron
as a cubic grid of magnetic domains $S_i$, whose north pole is either
pointing upward ($S_i=+1$) or downward ($S_i=-1)$. The external field
pushes on our domain with a force $H(t)$, which starts pointing down
($H(t=0)\ll 0$) and will increase with time.
Iron is magnetic because a domain has lower energy when it is magnetized
in the same direction as its neighbors; the force on site $i$ from the
six neighboring domains $S_j$ in our model is of strength $J \sum_j S_j$.
Finally, we model
the effects of impurities, randomness in the domain shapes, and 
other kinds of disorder by introducing a random field%
  \footnote{Most disorder in magnets is not microscopically well described
  by random fields, but better modeled using random
  anisotropy or random bonds which do not break time-reversal invariance.
  For models of hysteresis, time-reversal symmetry is already broken by the 
  external field, and all three types of randomness are probably in the
  same universality class. That is, random field models will have the same 
  statistical behavior as more realistic models, at least for large
  avalanches and long times (see section~\ref{sbsec:Universality}).}
$h_i$, different for each domain and chosen from a normal distribution
with standard deviation $R \times J$. The net force on $S_i$ is thus
\begin{equation}
F_i = H(t) + \sum_j J S_j + h_i,
\end{equation}
corresponding to the energy function or Hamiltonian
\begin{equation}
{\cal H} = -J \sum_{\langle ij\rangle} S_i S_j - \sum_i (H(t)+ h_i) S_i.
\end{equation}
This model is called the {\em random-field Ising model} (RFIM), and its thermal
equilibrium properties have historically been studied as an archetypal
disordered system with glassy
dynamics. Here we ignore temperature (specializing to magnets where the thermal
fluctuations are too small to de-pin the domain walls), and assume that
each domain $S_i$ reorients whenever its local field $F_i$ changes sign,
to minimize its energy.

In our model, there are two situations where a domain will flip. It may
be induced to flip directly by a change in the external field $H(t)$, 
or it may be triggered to flip by the flipping of one of its neighboring
domains. The first case corresponds to nucleating a new avalanche; the second
propagates the avalanche outward from the triggering domain, see 
Fig.~\ref{fig:BigAndSmallAvalanches}(b). We assume that the external
field $H(t)$ changes slowly enough that each avalanche finishes (even ones
which sweep over the entire magnet) before the field changes appreciably.

\begin{figure}[thb]
  \begin{center}
    \epsfxsize=\hsize
    \epsffile{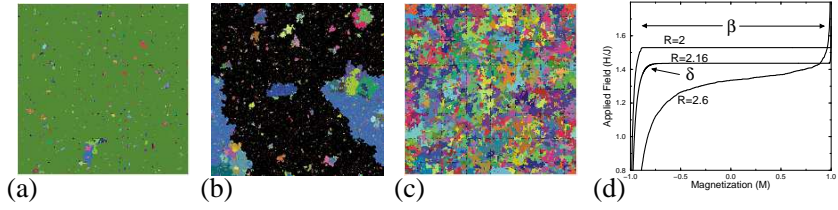}
  \end{center}
\caption{Phase transition in hysteresis and avalanche model;
(a)~one enormous avalanche (background) for small disorder $R<R_c$,
(b)~avalanches at all scales at $R=R_c$
(c)~many small avalanches for large disorder $R>R_c$.
(d)~The upper branch of the hysteresis loop develops a macroscopic jump 
in magnetization below $R_c$,
whose size diverges as $\Delta M \propto (R-R_c)^\beta$; at $R=R_c$ the
magnetization has a power-law singularity $M - M_c \sim (H-H_c)^{1/\delta}$. 
}
\label{fig:HysteresisTransition}
\end{figure}

These avalanches can become enormous, as in 
Fig.~\ref{fig:BigAndSmallAvalanches}(a).
How can an avalanche grow to over $10^5$ domains, but then halt? In our
model, it happens only when the disorder $R$ and the field $H(t)$ are
near a {\em critical point} (Fig.~\ref{fig:HysteresisTransition}).
For large disorder compared to the
coupling between domains $R\gg J \equiv 1$, each domain turns over 
almost independently (roughly
when the external field $H(t)$ cancels the local random field $h_i$, with
the alignment of the neighbors shifting the transition only slightly);
all avalanches will be 
tiny. For small disorder compared to the coupling $R\ll J$, there will
typically be one large avalanche that flips most of the domains
in the sample. (The external
field $H(t)$ needed to nucleate the first domain flip will be large, to 
counteract the field $-6 J$ from the unflipped neighbors without substantial
assistance from the random field; at this large external field most of the
neighboring domains will be triggered, \ldots eventually flipping most of the 
domains in the entire system). There is a critical disorder $R_c$ separating
two qualitative regimes of behavior---a phase $R>R_c$ where all avalanches
are small and a phase $R<R_c$ where one `infinite' avalanche flips a finite
fraction of all domains in the system (Fig.~\ref{fig:HysteresisTransition}).
Near%
  \footnote{One must also be near the field $H(t)=H_c$ where the infinite
  avalanche line ends.}
$R_c$, a growing avalanche doesn't
know whether it should grow forever or halt while small, so a distribution
of avalanches of all sizes might be expected--- giving us crackling noise.

\subsection{Real Barkhausen noise}
\label{sbsec:RealBarkhausen}

Is our model a correct description of Barkhausen noise in real magnets?
Our model does capture the qualitative physics rather well; near $R_c$
it exhibits crackling noise with a broad distribution of avalanche sizes
(Fig.~\ref{fig:BarkhausenIntro}(b)), and the individual
avalanches have the same kind of internal irregular dynamical structure as seen
in real Barkhausen avalanches (Fig.~\ref{fig:TimeSeriesPulses}).

\begin{figure}[thb]
  \begin{center}
    \epsfxsize=8cm
    \epsffile{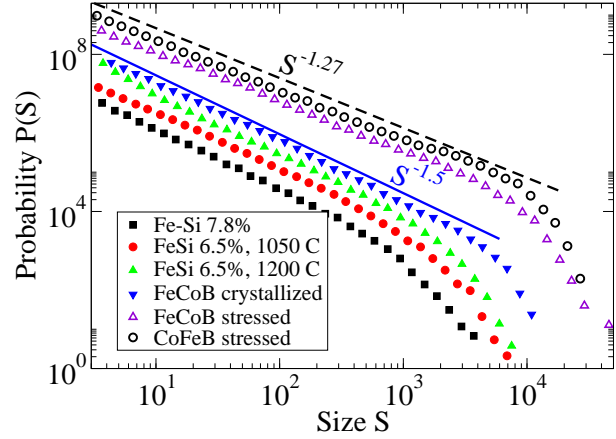}
  \end{center}
\caption{{\bf Avalanche size distributions} for magnetic Barkhausen noise in
various materials, from~\cite{DuriZapp00}. The materials fall into two
families, with power law exponent $D(S)\sim S^{-\tau}$ with $\tau \sim 1.5$
and $\tau\sim 1.27$. The cutoffs in the avalanche size distribution at large
$S$ are due to demagnetization effects, and scale as 
$S_{\mathrm{max}}~k^{-1/\sigma_k}$ (section~\ref{sbsec:PowerLaws}, 
Fig.~\ref{fig:AvalHisto}).
}
\label{fig:DuriZapp00Sizes}
\end{figure}

However, when examined in detail we find that there are problems with the 
model. A typical problem might be that the theoretically predicted power-law 
is wrong (Fig.~\ref{fig:DuriZapp00Sizes}). The probability density ${\cal D}(S)$
of an avalanche of size $S$ experimentally~\cite{DuriZapp00} 
decays as a power law 
$D(S) \sim S^{-\tau}$, with values of the exponent that cluster around
either $\tau=1.27$ or $\tau = 1.5$. In our model at $R_c$ and near $H_c$
we find $\tau = 1.6 \pm 0.06$, which is marginally compatible with the data
for the second group of samples (see Fig.~\ref{fig:Exponents}(a)).
If the power law were significantly different, our model would have
been ruled out.
We shall see in
section~\ref{sec:PowerLawsScalingFunctions}
that the power laws are key {\em universal} predictions of the theory, 
and if they do not agree with the experiment then the theory is
missing something crucial.

However, there are two other closely related avalanche models that 
describe the two groups of experiments well. One is a front propagation 
or fluid invasion model 
(which preceded ours~\cite{CiepRobb88,JiRobb92,Nara96,KoilRobb00}), also
usually written as a random-field Ising model but here starting with
an existing front and allowing new avalanches only contiguous to previously
flipped domains (water can invade new pores only next to currently wet pores).
The front-propagation model does well in describing the dynamics and size 
distribution of the materials
with $\tau \approx 1.27$. The other model can either be viewed as the
motion of a flat domain wall~\cite{AlesBeatBert+90e,AlesBeatBert+90t}
or the mean-field limit of the random-field Ising
model~\cite{CizeZappDuri+97,ZappCizeDuri+98}; it has a value $\tau=3/2$
nicely describing the other class of materials. These models
incorporate the long-range magnetic interactions between flipped domains,
which experimentally produce demagnetizing fields that 
induce the magnetic domain wall to move rather rigidly. 

\begin{figure}[thb]
  \begin{center}
    \epsfxsize=\hsize
    \epsffile{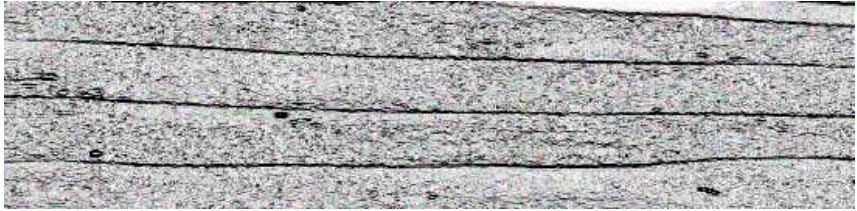}
  \end{center}
\caption{{\bf Domain structure in a real magnet}, measured with a scanning
electron microscope, digitally processed to detect edges. The horizontal
lines are magnetic domain walls. Courtesy of K.~Zaveta,
from~\cite{url:DurinBarkhausen}.
}
\label{fig:DurinRealDomainEdges}
\end{figure}

It is easy to measure the exponents of a power-law decay, but it is difficult
to measure them well. We could argue that $\tau=1.6$ is close enough to
some of the experimental measurements that our model is not ruled out. But
even a brief investigation of the qualitative behavior in these systems 
leads us to abandon hope. Fig.~\ref{fig:DurinRealDomainEdges} shows the
domain-wall structure in a real magnet. The domain walls are indeed flat,
and the way in which they advance (looking at animated versions of this
figure, or observing directly through a microscope) is not by nucleating new 
walls, but by motion of existing boundaries. 

\section{Why Crackling Noise?}
\label{sec:WhyCrackling}

In this section we introduce {\em universality}, the {\em renormalization
group}, and {\em self-similarity}. These three notions grew out of the
theory of continuous phase transitions in thermal statistical mechanics,
and are the basis of our understanding of a broad variety of systems, 
from quantum transitions (insulator to superconductor) to dynamical 
systems (the onset of chaos)~\cite[chapter 12]{Seth06}. We will illustrate
and explain these three topics not only with our model for crackling noise
in magnets, but also using classic problems in statistical 
mechanics---percolation and the liquid-gas transition. The 
renormalization group is our tool for understanding crackling noise.

\subsection{Universality}
\label{sbsec:Universality}

Consider holding a sheet of paper by one edge, and sequentially punching 
holes of a fixed size at random positions. If we punch out only a few holes,
the paper will remain intact, but if the density of holes approaches one
the paper will fall apart into small shreds. There is a phase transition
somewhere in between where the paper first ceases to hold together. This
transition is called {\em percolation}.

  \cfig{Universality in percolation}
  {. Universality suggests that the entire morphology of the percolation
  cluster at $p_c$ should be independent of microscopic details. On the
  top, we have bond percolation, where the bonds connecting nodes on a
  square lattice are occupied at random with probability $p$; the top
  right shows the infinite cluster on a $1024\times1024$ lattice
  at the critical point. On the bottom, we have site percolation on a
  triangular lattice, where it is the hexagonal sites that are
  occupied. Even though the microscopic lattices
  and occupation rules are completely different, the resulting clusters
  look statistically identical at their critical points.
  (One should note that the site percolation
  cluster is slightly less dark. Universality holds up to overall scale changes,
  here up to a change in the density.)
  }
  {PercolationUniversality}{0.99\hsize}
  {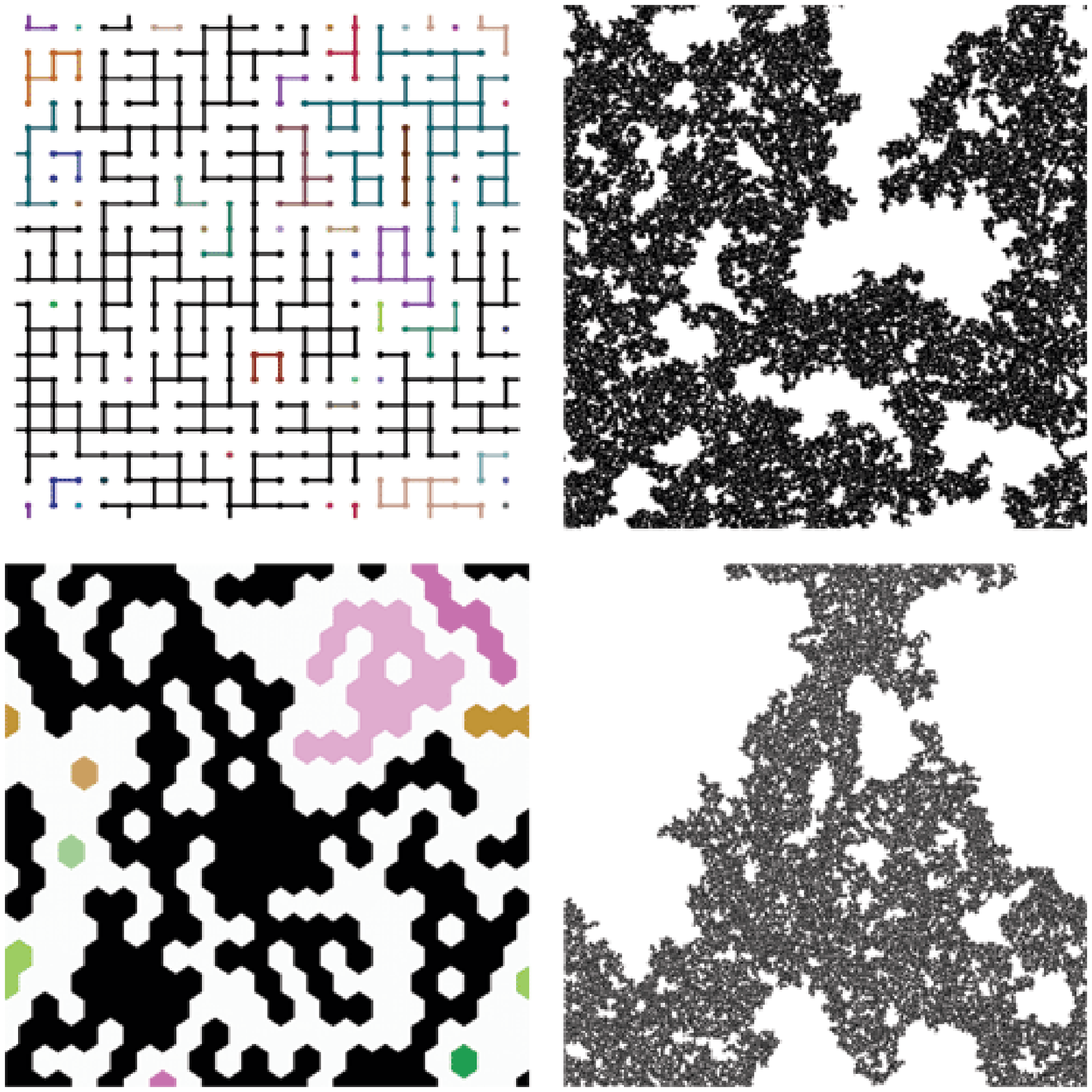}

Fig.~\ref{fig:PercolationUniversality} shows two somewhat different microscopic
realizations of this problem. On the top we see a square mesh of bonds,
where we remove all but a fraction $p$ of the bonds. On the bottom we
see a lattice of hexagonal regions punched out at random. Microscopically,
these two processes seem completely different (left figures). But 
if we observe large simulations and examine them at long length scales,
the two microscopic realizations yield statistically identical types of
percolation clusters (right Figs~\ref{fig:PercolationUniversality}).

{\em Universality} arises when the statistical morphology of a 
system at a phase transition is largely independent of the microscopic details 
of the system, depending only on the type (or {\em universality class}) of
the transition. This should not come as a complete surprise; we describe
most liquids with the same continuum laws (the Navier-Stokes equations)
despite their different molecular makeups, with only the mass density and
the viscosity depending on microscopic details. Similarly most 
solids obey elasticity theory on scales large compared to their constituent 
particles. For these phases we understand the material independence of the
constituent equations by taking a continuum
limit, assuming all behavior is slowly varying on the scale of the particles.
At critical points, because our system is rugged and 
complicated all the way down to the
lattice scale, we will need a more sophisticated way of understanding 
universality (section~\ref{sbsec:RenormalizationGroup}).

\cfig{Universality}
  {. (a)~Universality at the liquid--gas critical point.
  The liquid--gas coexistence lines ($\rho(T)/\rho_c$ versus $T/T_c$)
  for a variety of atoms and small molecules, near their critical points
  $(T_c, \rho_c)$~\cite{Gugg45}. The gas phases lie on the upper left;
  the liquid phase region is to the upper right; densities in between
  will separate into a portion each of coexisting liquid and gas.
  The curve is a fit to the argon data,
  $\rho/\rho_c = 1 + s (1-T/T_c) \pm \rho_0 (1-T/T_c)^\beta$
  with $s=0.75$, $\rho_0 = 1.75$, and $\beta=1/3$~\cite{Gugg45}.
  (b)~Universality: ferromagnetic--paramagnetic critical point.
  Magnetization versus temperature for a uniaxial antiferromagnet
  MnF$_2$~\cite{HellBene62}. The tilted line in~(a) corresponds to the
  vertical axis in~(b).}
  {Universality}{0.8\hsize}
  {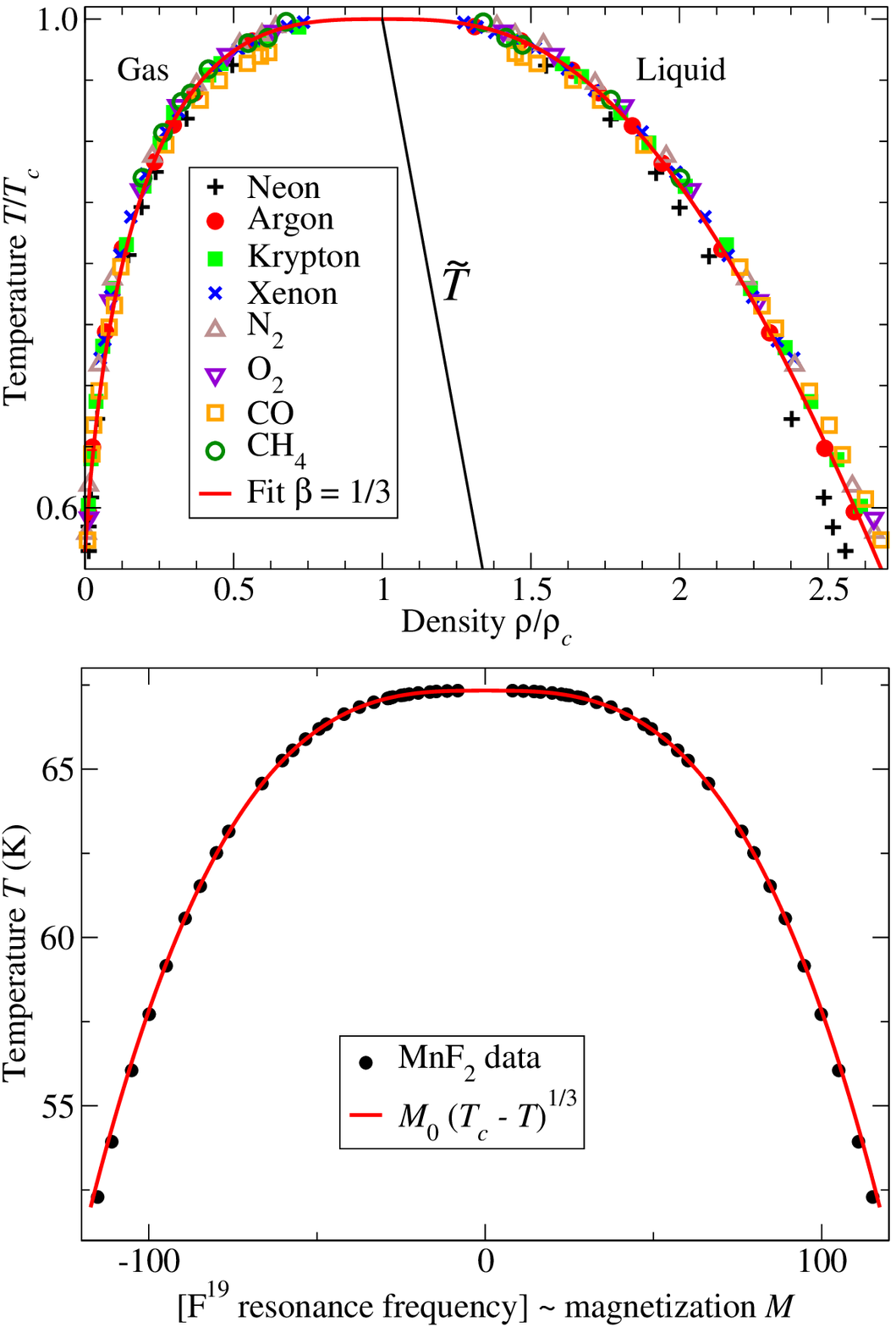}

Most thermal phase transitions are not continuous, and do not have
structure on all scales. Water is water until it is heated past the boiling
point, after which it abruptly turns to vapor 
(with a big drop in density $\rho$). Even boiling water, though,
can show big fluctuations at high enough pressure.
As we raise the pressure
the boiling temperature gets larger and the density drop gets smaller,
until at a certain pressure and temperature $(P_c,T_c)$ the two densities
become equal (to $\rho_c$).
At this critical point the H$_2$O molecules do not know which phase
they want to be in, and they show fluctuations on all scales.

Figure~\ref{fig:Universality}(a) shows that the approach to this critical
point has universal features. If we divide the density by
$\rho_c$ and the temperature by $T_c$, many different liquids share the
same $\rho,T$ phase diagram, 
\begin{equation}
\rho^{\mathrm{Ar}}(T) = A \rho^{\mathrm{CO}}(BT).
\end{equation}
This is partly for mundane reasons; most of the
molecules are roughly spherical, and have weak interactions. 

But figure~\ref{fig:Universality}(b) shows the phase diagram of a completely 
different system, a magnet whose magnetization $M(T)$ vanishes as the 
temperature
is raised past its $T_c$. The curve here does not agree with the curves
in the liquid-gas collapse, but it can agree if we shear the
axes slightly:
\begin{equation}
\label{eq:MrhoCoordChange}
\rho^{\mathrm{Ar}}(T) = A_1 M(BT) + A_2 + A_3 T.
\end{equation}
Nature has decided that the `true' natural vertical coordinate for 
the liquid-gas transition is not $T$ but the line $T=(\rho-A_2)/A_3$.
Careful experiments, measuring a wide variety of
quantities, show that the liquid-gas transition and this type of magnet
share many properties, except for smooth changes of coordinates like
that in eq~\ref{eq:MrhoCoordChange}. Indeed, they share these properties
also with a theoretical model---the (thermal, non-random, 
three-dimensional) Ising model. Universality not only tied disparate
experiments together, it also allows our theories to work.

The agreement between Figs~\ref{fig:Universality}(a) and (b) may not seem
so exciting; both are pretty smooth curves. But notice that they do not
have parabolic tops (as one would expect from the maxima of a typical function).
Careful measurements show that the magnet, the liquid--gas critical point,
and the Ising model all vary as $(1-T/T_c)^\beta$ with the same, probably
irrational exponent $\beta$; the best theoretical estimates have
$\beta = 0.325\pm 0.005$~\protect\cite[chapter 28]{Zinn96}.
This characteristic power law represents the effects of the large
fluctuations at the critical points (the peaks of these graphs).
Like the avalanche size distribution exponent $\tau$, $\beta$ is a 
{\em universal critical exponent} (section~\ref{sbsec:PowerLaws}).

\subsection{Renormalization group}
\label{sbsec:RenormalizationGroup}

\begin{figure}[thb]
  \begin{center}
    \epsfxsize=\hsize
    \epsffile{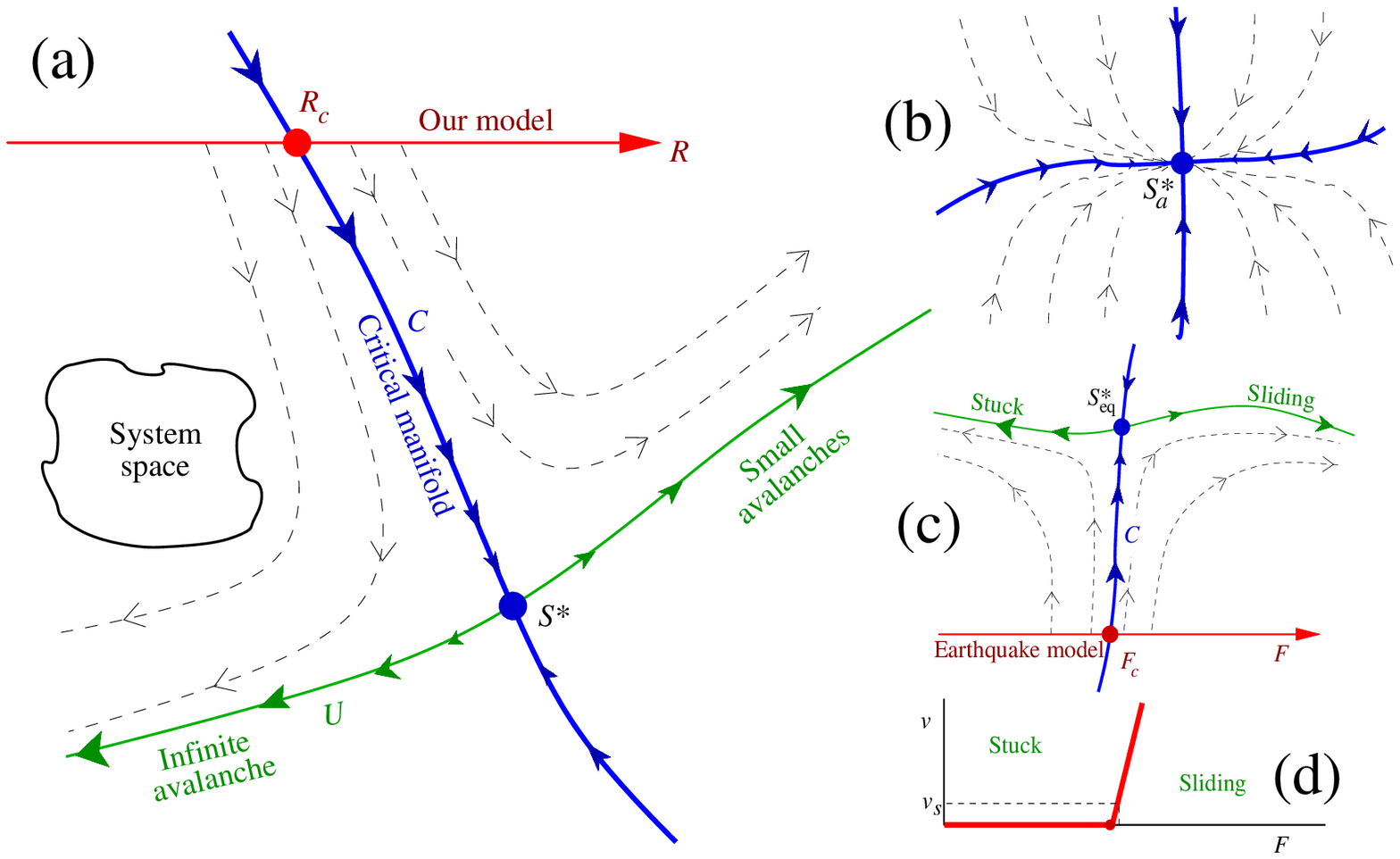}
  \end{center}
\caption{{\bf Renormalization--group flows}~\cite{SethDahmMyer01}. 
(a)~{\bf Flows describing a phase transition.} The
renormalization--group uses coarse--graining to longer length scales to
produce a mapping from the space of physical systems into itself. Consider
the space of all possible systems exhibiting magnetic hysteresis (including,
both real models and experimental systems). Each
model can be coarse--grained, removing some fraction of the
microscopic domains and introducing more complex dynamical rules so
that the remaining domains still flip over at the same external fields.
This defines a mapping of our space of models into itself. A fixed point
$\mathbf{S^*}$ in this space will be self--similar: because it maps
into itself upon coarse--graining, it must have the same behavior on
different length scales. Points that flow into $\mathbf{S^*}$ under
coarse--graining share this self--similar behavior on sufficiently long
length scales: they all share the same {\em universality class}.
(b)~{\bf Attracting fixed point}~\cite{SethDahmMyer01}.
Often there will be fixed points that
attract in all directions. These fixed points describe typical behavior:
phases rather than phase transitions. 
Most phases are rather boring on long length scales. In more interesting
cases, like random walks, systems can exhibit self-similarity and 
power laws without special tuning of parameters. This is called
{\em generic scale invariance}.
(c,d)~{\bf Flows for a front--propagation model}~\cite{SethDahmMyer01}.
The front propagation model has a critical field $F_c$ at which the
front changes from a pinned to sliding state.
(c)~Coarse-graining defines a flow on the space of earthquake models.
The critical manifold $\mathbf{C}$, consisting of
models which flow into $\mathbf{S^*}_{fp}$, separating stuck faults
from faults which slide forward with an average velocity $v(F)$.
(d)~The velocity varies with the external force $F$ across
the fault as a power law
$v(F)\sim (F-F_c)^\beta$. Clever experiments, or long--range fields, can
act to control not the external field, but the position: changing
the front displacement slowly sets $v\approx 0$, thus self--tuning 
$F\approx F_c$.  This is one example of {\em self-organized criticality}.
}
\label{fig:RGFlowCombined}
\end{figure}

Our explanation for universality, and our theoretical framework for 
studying crackling noise, is the {\em renormalization group}.
The renormalization group starts with a remarkable abstraction: it works
in an enormous `system space'. Different points in system space represent
different materials under different experimental conditions, or different
theoretical models with different interactions and evolution rules. 
For example, in Fig.~\ref{fig:RGFlowCombined}(a) we consider the space
of all possible models and experiments on hysteresis and avalanches,
with a different dimension
for each possible parameter (disorder $R$,
coupling $J$, next-neighbor coupling, \ldots) and for each parameter
in an experiment (chemical composition, annealing time, \ldots). Our
theoretical model will traverse a line in this infinite-dimensional space
as the disorder $R$ is varied.

\cfig{Ising model at $\mathbf{T_c}$: coarse-graining}
  {. Coarse-graining of a snapshot of the two-dimensional Ising
  model at its critical point. Each
  coarse-graining operation
  changes the length scale by a factor $B=3$.
  Each coarse-grained spin points in the
  direction given by the majority of the nine fine-grained spins it replaces.
  This type of coarse-graining is the basic operation of the real-space
  renormalization group.
  }
  {IsingCoarseGraining}{0.99\hsize}{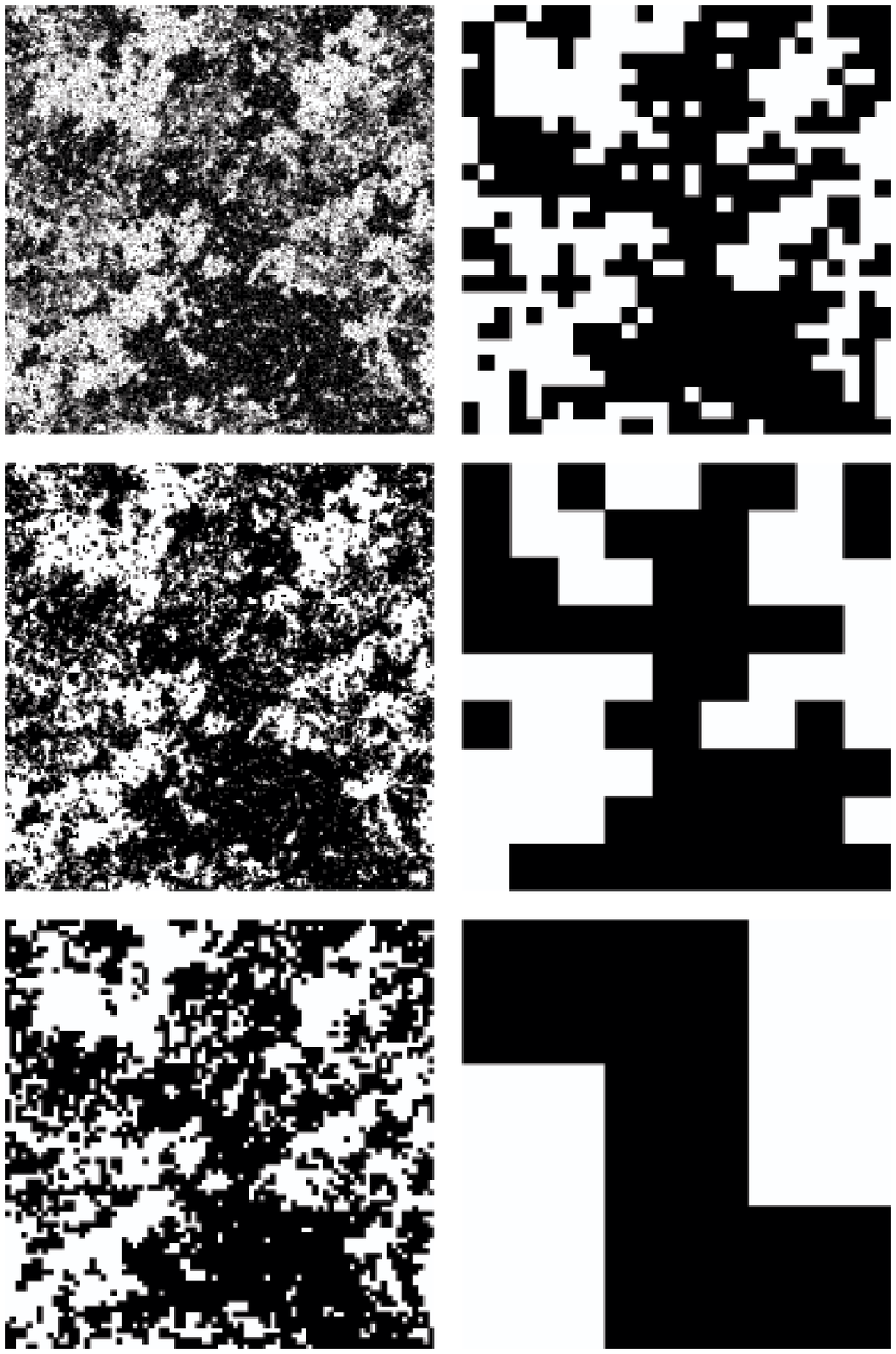}

The renormalization group studies the way in which system space maps
into itself under {\em coarse-graining}. The coarse-graining operation
shrinks the system and removes microscopic degrees of freedom. Ignoring
the microscopic degrees of freedom yields a new physical system with
the same properties at long length scales, but with different (renormalized)
values of the parameters. As an example, figure~\ref{fig:IsingCoarseGraining}
shows a real-space renormalization-group `majority rule' coarse-graining
procedure applied to the Ising model. Many approximate methods have
been devised to implement this coarse-graining operation (real-space,
$\epsilon$-expansions, Monte Carlo\ldots) which we will not discuss here.

Under coarse-graining, we often find a fixed point $S^*$ in system space.
All of the systems that flow into this fixed point under coarse-graining
will share the same long-wavelength properties, and will hence be in
the same universality class. 

Fig.~\ref{fig:RGFlowCombined}(a) shows the case of a fixed point $S^*$ with 
one unstable direction. Points deviating from $S^*$ in this direction
will flow away from it under coarse-graining. There is a surface $C$ of
points which do flow into the fixed point, which separates system space into
two different phases (say, one with all small avalanches and one with
one system-spanning, `infinite' avalanche). The set $C$ represents
a universality class of systems at their critical points. Thus,
fixed points with one
unstable direction represent phase transitions. 

Cases like the liquid-gas transition with two tuning parameters 
$(T_c,\rho_c)$ determining the critical point will have two unstable
directions. What happens when we have no unstable directions? The fixed-point
$S_a$ in Fig.~\ref{fig:RGFlowCombined}(b) represents an entire region
in system space sharing the same long-wavelength properties; it represents
a {\em phase} of the system. Usually phases do not show structure on all
scales, but some cases (like random walks) show this {\em generic scale
invariance}.

Sometimes the external conditions acting on a system naturally drive it
to stay at or near a critical point, allowing one to spontaneously
observe fluctuations on
\forcePageBreakNoSkip
all scales. A good example is provided by
certain models of earthquake fault dynamics. Fig.~\ref{fig:RGFlowCombined}(c)
schematically shows the renormalization-group flow for an earthquake model.
It has a {\em depinning} fixed point representing the external force
$F$ across the fault at which the fault begins to slip; for $F>F_c$ the
fault will slide with a velocity $v(F)\sim (F-F_c)^\beta$. Only near
$F_c$ will the system exhibit 
earthquakes of all sizes. Continental plates,
however, do not impose constant forces on the faults between them; they
impose a rather small constant velocity (on the order of centimeters per year,
much slower than the typical fault speed during an earthquake). 
As illustrated in Fig.~\ref{fig:RGFlowCombined}(d), this naturally sets
the fault almost precisely at its critical point, tuning the system to
its phase transition. This is called {\em self-organized criticality}.

In broad, the existence of fixed points under renormalization group
coarse-graining is our fundamental explanation for universality. 
Real magnets, Ising models, and liquid-gas systems all apparently flow
to the same fixed points under coarse graining, and hence show the
same characteristic fluctuations and behaviors at long length scales.

\section{Self-similarity and its consequences}
\label{sec:PowerLawsScalingFunctions}

 \cfig{Avalanches: scale invariance}
  {. Magnifications of a cross-section of all the avalanches in a run of our
  hysteresis model
  each one the lower right-hand quarter of the previous. The system started
  with a billion domains (1000$^3$). Each avalanche is
  shown in a different shade. Again, the larger scales look statistically the
  same.
  \Index{Critical point!emergent scale invariance!avalanche $R_c$}
  \Index{Scale invariance!avalanche model $R_c$}
  \Index{Hysteresis and avalanches!scale invariance at $R_c$}
  \Index{Avalanche!and hysteresis!scale invariance at $R_c$}
  \Index{Barkhausen noise, magnetic!scale invariance at $R_c$}
  }
  {AvalanchesCrossSectionScaling}
  {0.99\hsize}{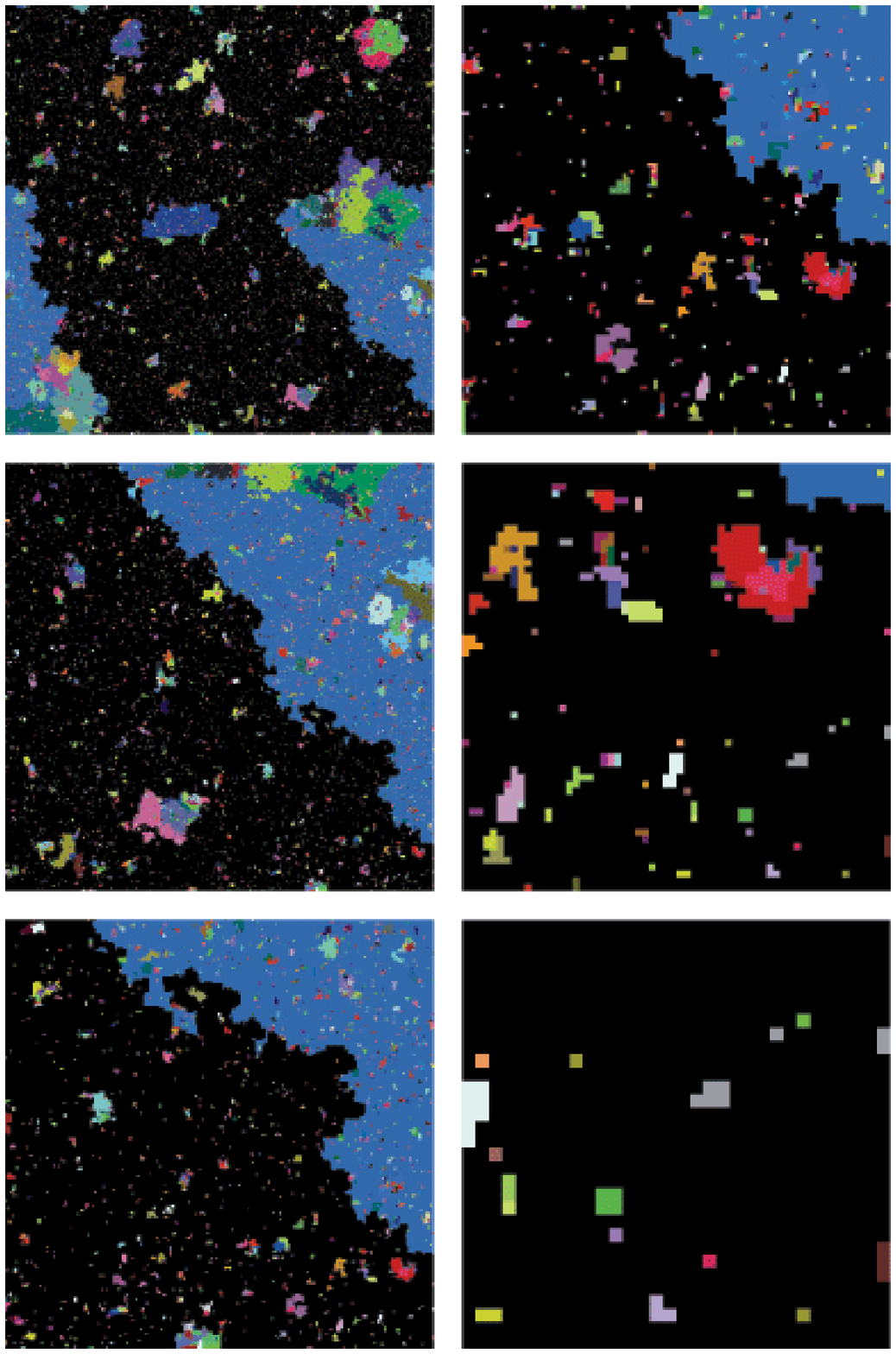}

The most striking feature of crackling noise and other critical systems
is self-similarity, or scale invariance. We can see this vividly in the
patterns observed in the Ising model (upper left, 
Fig.~\ref{fig:IsingCoarseGraining}), percolation 
(Fig.~\ref{fig:PercolationUniversality}), and our RFIM for crackling noise
(Fig.~\ref{fig:BigAndSmallAvalanches}a). Each shows roughness,
irregularities, and holes on all scales at the critical point. This roughness
and fractal-looking structure stems at root from a hidden symmetry in
the problem: these systems are (statistically) invariant under a change
in {\em length scale}.

Consider Fig.~\ref{fig:AvalanchesCrossSectionScaling}, depicting the 
self-similarity of the avalanches in our RFIM simulation at the critical
disorder. The upper-right figure shows the entire system, and each
succeeding picture zooms in by another factor of two. Each zoomed-in
picture has a black `background' showing the largest avalanche spanning
the entire system, and a variety of smaller avalanches of various sizes.
If you blur your eyes a bit, the figures should look roughly alike. This
rescaling and blurring process is the renormalization-group coarse-graining
transformation. 

How does the renormalization group explain self-similarity? The
fixed point $S^*$ under the renormalization group is the same after
coarse-graining (that's what it means to be a fixed point). Any other
system that flows to $S^*$ under coarse graining will also look self-similar
(except on the microscopic scales that are removed in the first few
steps of coarse-graining, during the flow to $S^*$). Hence systems
at their critical points naturally exhibit self-similarity.

This scale invariance can be thought of as an emergent symmetry, invariance
under changes of length scale. In a system invariant under a translational
symmetry,
the expectation of any function of two positions $x_1$ and $x_2$ can
be written in terms of the separation between the two points,
$\langle g(x_1, x_2)\rangle = {\cal G}(x_2-x_1)$. In just the same way, 
scale invariance will allow us to write functions of $N$ variables in
terms of {\em scaling functions} of $N-1$ variables---except that these
scaling functions are typically multiplied by power laws in one of these
variables.

\subsection{Power laws}
\label{sbsec:PowerLaws}

Let us begin with the case of functions of one variable. Consider the
avalanche size distribution $D(S)$ for a model, say the real
earthquakes in Fig.~\ref{fig:Earthquakes}(a), or our model for hysteresis
at its critical point. Imagine taking the same system, but increasing
the units of length with which we measure the system---stepping back,
blurring our eyes, and looking at the system on a coarse-grained level.
Let us multiply the spacing between markings on our rulers by a small
amount $B=(1+\epsilon)$. After coarsening, any length scales in the
problem (like the spatial extent $L$ of a particular avalanche) will
be divided by $B$. The avalanche size (volume) $S$ after coarse-graining
will also be smaller by some factor%
  \footnote{If the size of the avalanche were the cube of its length,
  then $c$ would equal three since 
  $(1+\epsilon)^{~3}=1+3\epsilon+O(\epsilon^2)$.
  In general, $c$ is the {\em fractal dimension} of the avalanche.}
$C = (1+c\epsilon)$. Finally, the overall scale of $D(S)$ after coarse-graining
will be changed by some factor $A=1+a\epsilon$.%
  \footnote{The same avalanches occur independent of your measuring
  instruments! But the probability $D(S)$ changes, because the {\em fraction}
  of large avalanches depends upon how many small avalanches you measure,
  and because the fraction per unit $S$ changes as the scale of $S$ changes.}
Hence under the coarse-graining we have
\begin{align}
\label{eq:DRescaleAtRc}
L' = L/B = L/(1+\epsilon),\nonumber\\
S' = S/C = S/(1+c\epsilon),\\
D' = A D = D (1+a\epsilon).\nonumber
\end{align}
Now the probability that the coarse-grained system has an avalanche of 
size $S'$ is given by the rescaled probability that the original system
had an avalanche of size $S=(1+c\epsilon) S'$:
\begin{equation}
D'(S') = A D(C S') = (1+a\epsilon) D\left((1+c \epsilon) S')\right).
\end{equation}
Here $D'(S')$ is the distribution measured with the new ruler: a smaller
avalanche with a larger probability density. Because we are at a 
self-similar critical point, the coarse-grained distribution $D'(S')$ 
should equal $D(S')$. Making $\epsilon$ infinitesimal leads us to a
differential equation:
\begin{align}
D(S') &= D'(S') = (1+a\epsilon) D\big((1+c\epsilon) S'\big), \nonumber\\
0 &= a \epsilon D + c \epsilon S' \frac{\d{D}}{\d{S}}, \nonumber\\
\frac{\d{D}}{\d{S}} &= -\frac{a D}{cS},
\end{align}
which has the general solution
\begin{equation}
\label{eq:DofSPowerLaw}
D = D_0 S^{-a/c}.
\end{equation}

Because the properties shared in a universality class only hold up to
overall scales, the constant $D_0$ is system dependent.
However, the exponents $a$, $c$, and $a/c$ are {\em universal}---independent
of experiment (within the universality class).
Some of these exponents have standard names:
the exponent $c$ giving the fractal dimension of the avalanche is usually
called $d_f$ or $1/\sigma\nu$.
The exponent $a/c$ giving the size
distribution law is called $\tau$ in percolation and in
most models of avalanches in magnets%
  \footnote{\label{note:taubar}
  In our RFIM for hysteresis, we use $\tau$ to denote the
  avalanche size law at the critical field and disorder 
  ($D(S,R_c,H_c)\sim S^{-\tau}$); integrated over the
  hysteresis loop $D_{\mathrm{int}} \propto S^{-\bar{\tau}}$ with
  $\bar{\tau} = \tau + \sigma\beta\delta$.
  }
and is related to the Gutenberg--Richter exponent for earthquakes%
  \footnote{We must not pretend that we have found the final
  explanation for the Gutenberg--Richter
  law. There are many different models that give exponents $\approx2/3$, but
  it remains controversial which of these, if any, are correct for
  real-world earthquakes.
  }
(Fig.~\ref{fig:Earthquakes}(b)).
Most measured quantities depending on one variable will have similar
power-law singularities at the critical point. For example, the 
distribution of avalanche durations and peak heights also have power-law forms.
This is because power laws are the only self-similar functions. If
$f(x) = x^{-\alpha}$, then on a new scale multiplying $x$ by $B$,
$f(B x) = B^{-\alpha} x^{-\alpha} \propto f(x)$. 

\begin{figure}[thb]
  \begin{center}
    \epsfxsize=\hsize
    \epsffile{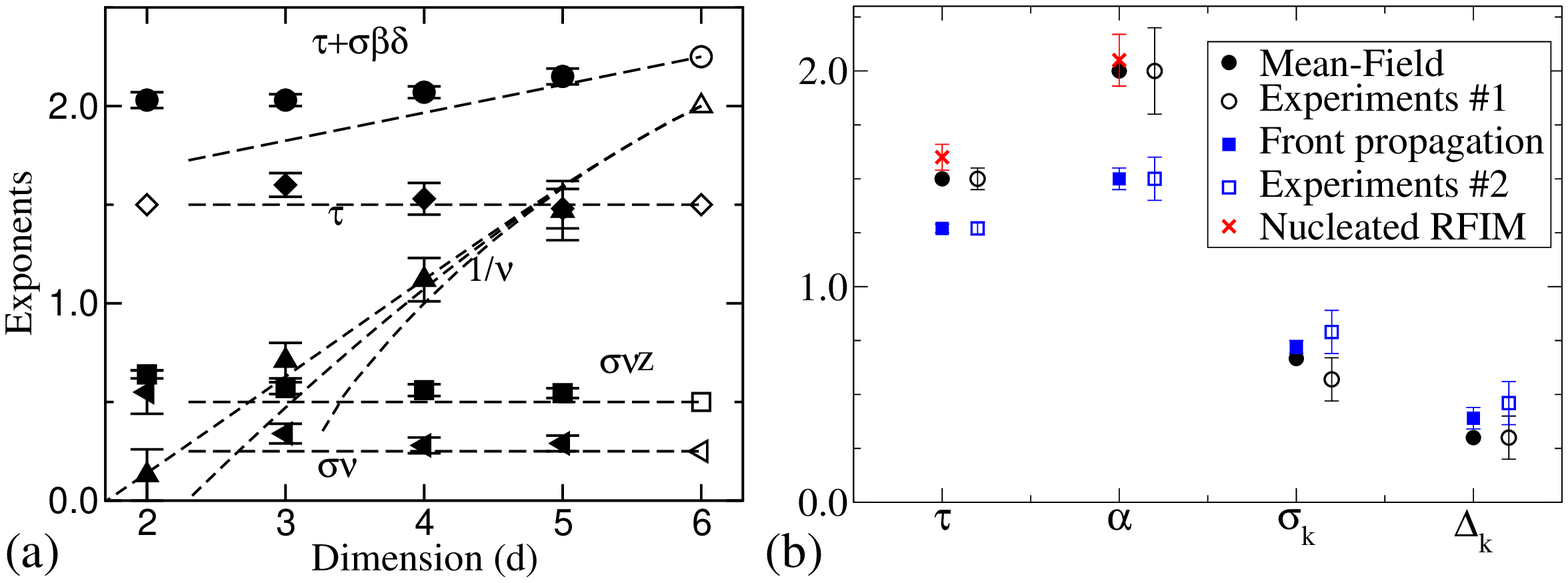}
  \end{center}
\caption{
(a)~{\bf Universal Critical Exponents in Various Spatial
Dimensions.}~\cite{SethDahmMyer01}
We test our $\epsilon$-expansion predictions~\cite{DahmSeth96} by
measuring the various critical exponents numerically
in up to five spatial dimensions~\cite{PerkDahmSeth95,PerkDahmSeth99}.
The various exponents are
described in the text. All of the exponents are calculated only to
linear order in $\epsilon$, except for the correlation length exponent $\nu$,
where we use results from other models. The agreement even in three
dimensions is remarkably good, considering that we're expanding in $6-D$ where
$D=3$!
(b)~{\bf Universal Critical Exponents vs.\
Experiment}~\cite{DuriZapp00}. The exponent
$\tau$ gives the probability density $\sim S^{-\tau}$ of having an
avalanche of size $S$, $\alpha$ gives the probability
density $\sim T^{-\alpha}$ of having an avalanche of duration $T$, 
and the exponents $\sigma_k$ and $\Delta_k$ describe the cutoff in the
avalanche sizes and durations, respectively, due to the demagnetizing field
(see~\cite{DuriZapp00}). The experimental samples group naturally into two
groups, described by the mean-field and front-propagation universality
classes; none appear to be described well by our nucleated RFIM.
}
\label{fig:Exponents}
\end{figure}

Crackling noise involves several power laws.
We've seen that the probability of having an avalanche of size $S$ goes as 
$S^{-\tau}$. 
In our RFIM model for hysteresis, if one is at a distance $R-R_c$ from
the critical point, there will be a cutoff in the avalanche size distribution.
The typical largest spatial extent $L$ of an avalanche is called the
{\em correlation length} $\xi$, which scales as $\xi\sim (R-R_c)^{-\nu}$.
The cutoff in the avalanche size $S$ scales as $(R-R_c)^{-\sigma}$
(Fig~\ref{fig:AvalHisto}).
In other models, demagnetization effects (parameterized by $k$)
lead to cutoffs in the avalanche size distribution~\cite{DuriZapp00},
with analogous critical exponents $\nu_k$ and $\sigma_k$.
The size and spatial extent of the avalanches are related to one
another by a power law $S \sim L^{d_f}$, where $d_f = 1/\sigma \nu$ 
is called the {\em fractal dimension}. The duration of an 
avalanche goes as $L^z$.
The probability of having an avalanche of duration $T$ goes as 
$T^{-\alpha}$, where $\alpha = (\tau-1)/\sigma\nu z +1$.%
  \footnote{Notice that we can write $d_f$ and $\alpha$ in terms of
  the other exponents. These are {\em exponent relations}; all of the
  exponents can typically be written in terms of two or three basic ones.
  We shall derive some exponent relations in 
  section~\ref{sbsec:ScalingFunctions}.}
In our RFIM, the jump in magnetization
goes as $(R-R_c)^\beta$, and at $R_c$ the magnetization 
$(M-M_c) \sim (H-H_c)^{1/\delta}$ (Fig.~\ref{fig:HysteresisTransition}(d)).

To specialists in critical phenomena, these exponents are central; whole
conversations often rotate around various combinations of Greek letters.
We know how to calculate critical exponents from various analytical
approaches; given an implementation of the renormalization group
they can be derived from the eigenvalues of the linearization
of the renormalization-group flow around the fixed-point $S^*$ in
Fig.~\ref{fig:RGFlowCombined}. Figure~\ref{fig:Exponents}(a) shows our numerical
estimates for several critical exponents of the RFIM model for Barkhausen
noise~\cite{PerkDahmSeth95,PerkDahmSeth99}, together with our $6-\epsilon$
expansions results~\cite{DahmSeth93,DahmSeth96}. Of course, the
challenge is not to get analytical work to agree with numerics: it is
to get theory to agree with experiment.
Figure~\ref{fig:Exponents}(b) compares recent Barkhausen experimental 
measurements of the critical exponents to values from the three 
theoretical models.

\subsection{Scaling functions}
\label{sbsec:ScalingFunctions}

Critical exponents are not everything, however. Many other
scaling predictions are easily extracted from numerical simulations, even
if they are inconvenient to calculate analytically.
(Universality should extend even to those properties that we have
not been able to write formul{\ae} for.) In particular, there
are an abundance of functions of two or more variables that one can
measure, which are predicted to take universal {\em scaling forms}.

\subsubsection{Average pulse shape}
\label{sbsbsec:PulseShape}

\begin{figure}[thb]
  \begin{center}
    \epsfxsize=\hsize
    \epsffile{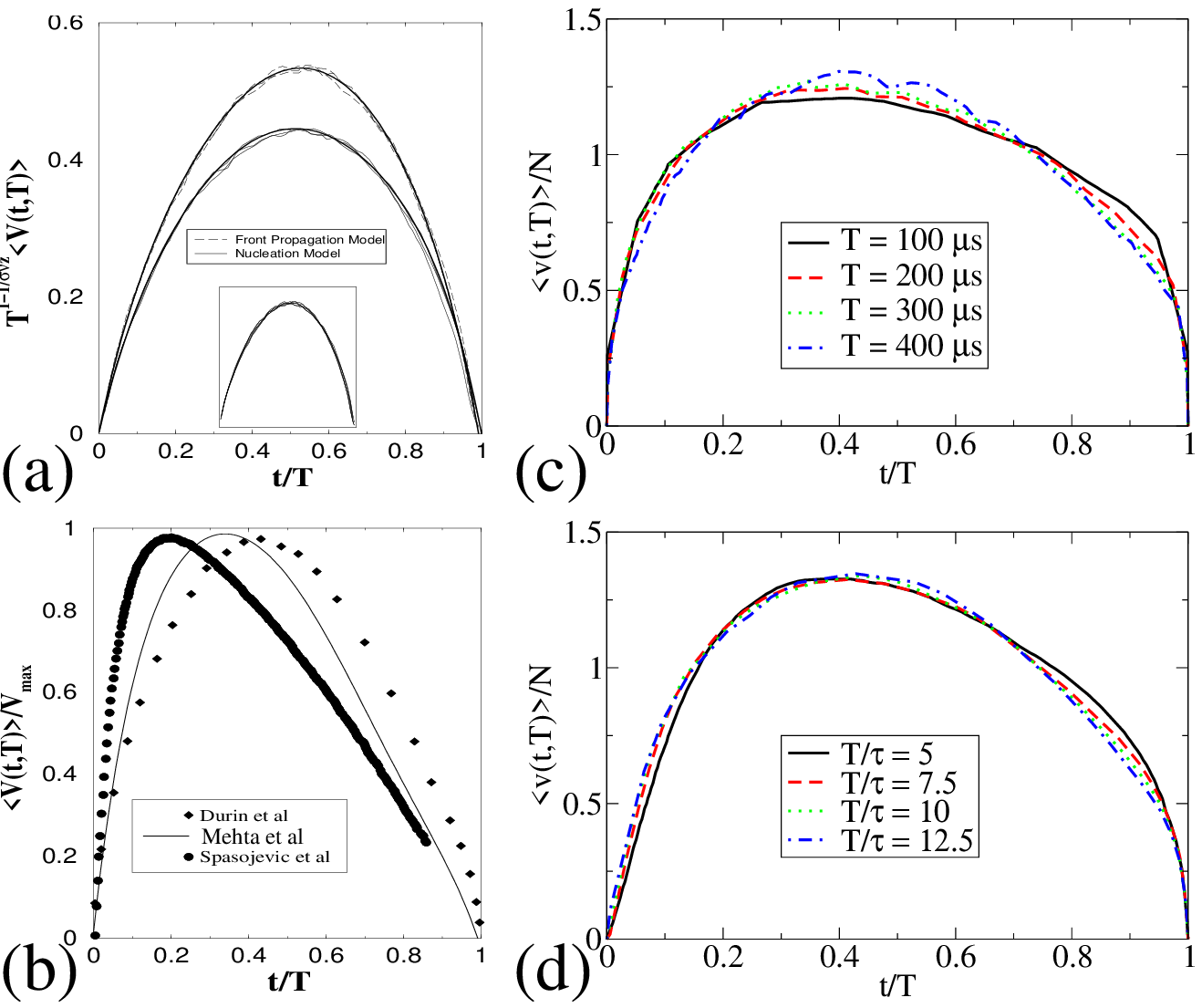}
  \end{center}
\caption{
(a)~{\bf Theoretical average pulse shape scaling functions}
for our nucleated model and the front propagation model~\cite{MehtMillDahm+02}.
The overall height is non-universal; the two curves are otherwise
extremely similar.
The front propagation model has $1/\sigma\nu z = 1.72\pm0.03$ in this
collapse; our nucleation model has $1/\sigma\nu z = 1.75\pm 0.03$ 
(there is no reason to believe these two should agree). The
inset shows the two curves rescaled to the same height (the overall
height is a non--universal feature): they are quantitatively different,
but far more similar to one another than either is to the experimental
curves in part~(b). 
(b)~{\bf Comparison of experimental average pulse shapes} for fixed pulse
duration, as measured by three different
groups~\cite{MehtMillDahm+02,DuriZapp00,DuriZapp01,SpasBukvMilo+96}.
Notice that both theory curves are much more symmetric than those of the
experiments. Notice also that the three experiments do not agree.
This was a serious challenge to our ideas about the universality of the
dynamics of crackling noise~\cite{SethDahmMyer01}.
(c)~{\bf Pulse shape asymmetry experiment}~\cite{ZappCastCola05}.
Careful experiments show a weak but systematic
duration dependence in the collapse of the
average Barkhausen pulse shape. The longer pulses (larger avalanches) are
systematically more symmetric (approaching the theoretical prediction).
(d)~{\bf Pulse shape asymmetry theory}~\cite{ZappCastCola05}.
Incorporating the time-retardation effects of eddy currents into 
the theoretical model produces a similar systematic effect. The non-universal
effects of eddy currents are in principle irrelevant for extremely large
avalanches.
}
\label{fig:PulseShapeCombined}
\end{figure}

For example, let us consider the time history $V(t)$ of the avalanches
(Fig.~\ref{fig:TimeSeriesPulses}). Each avalanche has large fluctuations,
but one can average over many avalanches to get a typical shape. 
Averaging $V(t)$ together for large and small avalanches would seem silly,
since only the large avalanches will last long enough to contribute at
late times. The experimentalists originally
addressed this issue~\cite{SpasBukvMilo+96} by rescaling the voltages of
each avalanche by the peak voltage and the time by the duration,
and then averaging these rescaled curves
(dark circles in Fig.~\ref{fig:PulseShapeCombined}b).
But there are so many more small avalanches than large ones that it seems
more sensible to study them separately.

Consider the average voltage $\bar V(T,t)$ over all avalanches of duration $T$.
Universality suggests that this average should be the same for all experiments
and (correct) theories, apart from an overall shift in time and voltage
scales:
\begin{equation}
\bar V_{\mathrm{exp}}(T,t) = A \bar V_{\mathrm{th}}(T/B, t/B).
\end{equation}
Comparing systems with a shifted time scale becomes simpler if we change
variables; let $v(T, t/T) = \bar V(T,t)$. Now, as we did for the avalanche
size distribution in section~\ref{sbsec:PowerLaws}, let us compare a system
with itself, after coarse-graining by a small factor $B=1/(1-\epsilon)$:
\begin{equation}
v(T,t/T) = A v(T/B, t/T) = (1+a) v\left((1-\epsilon)T, t/T\right).
\end{equation}
Again, making $\epsilon$ small we find $a v = T\, \partial v/\partial T$,
with solution $v(T,t/T) = v_0(T) T^a$. Here the integration constant $v_0$
will now depend on $t/T$, so we arrive at the scaling form
\begin{equation}
\label{eq:PulseShapeScaling}
\bar V(T,t) = T^a {\cal V}(t/T)
\end{equation}
where the scaling function ${\cal V} \equiv v_0$ (up to an overall constant
factor) is a universal prediction of the theory.

Can we write the exponent $a$ in terms of the exponents we already know?
Since the size of an avalanche is defined as the integral of $V(t)$,
we can use
\forcePageBreak
the scaling relation (eq~\ref{eq:PulseShapeScaling}) to write
an expression for the average size of an avalanche of duration $T$,
\begin{equation}
\bar S(T) = \int \bar V(t,T)\,\d{t} 
	= \int T^a {\cal V}(t/T)\,\d{t} \sim T^{a+1}.
\end{equation}
We also know that avalanches have fractal dimension $d_f=1/\sigma \nu$,
so $S \sim L^{1/\sigma \nu}$, and that the duration of an avalanche
of size $L$ goes as $T\sim L^z$. Hence 
$\bar S(T) \sim T^{1/\sigma \nu z}\sim T^{a+1}$, and $a = 1/\sigma \nu z -1$.
This is an example of an {\em exponent relation}.

Can we use the experimental pulse shape to figure out which theory is 
correct? Fig.~\ref{fig:PulseShapeCombined}(a) shows a {\em scaling collapse}
of the pulse shapes for our RFIM and the front propagation model.
The scaling collapse tests the scaling form eq~\ref{eq:PulseShapeScaling}
by attempting to plot the scaling function ${\cal V}(t/T)=T^{-a} V(t,T)$
for multiple durations $T$ on the same plot.
The two theoretical pulse shapes look remarkably similar to one another,
and almost perfectly time-reversal symmetric. The mean-field model also
is time-reversal symmetric.%
  \footnote{The mean--field model apparently has a scaling function which
  is a perfect inverted parabola~\cite{Kunt01}. The (otherwise 
  similar) rigid-domain-wall model, interestingly, has a different average
  pulse shape, that of one lobe of a sinusoid~\cite[eq. 3.102]{DuriZapp04}.}
Thus none of the
theories describe the strongly skewed experimental data 
(Fig.~\ref{fig:PulseShapeCombined}). Indeed, the experiments did not even
agree with one another, calling into question whether universality holds
for the dynamical properties.

This was recognized as a serious challenge to our whole theoretical
picture~\cite{KuntSeth00,SethDahmMyer01,MehtMillDahm+02}. An elegant,
convincing physical explanation was developed by Colaiori 
{\em et al.}~\cite{ColaAlavDuri04}, who attribute the asymmetry to eddy
currents, whose slow decay lead to a time-dependent damping of the 
domain wall mobility. Incorporating these eddy current effects
into the model leads to a clear correspondence between their eddy-current
theory and experiment, see Fig.~\ref{fig:PulseShapeCombined}(c,d). 
In those figures,
notice that the scaling `collapses' are imperfect, becoming more symmetric
for avalanches of longer durations. These eddy-current effects are 
theoretically {\em irrelevant} perturbations; under coarse-graining,
they disappear. So, the original models are in principle correct,%
  \footnote{One might wonder how the original models do so well for
  the avalanche size distributions and other properties, when they 
  have such problems with the pulse shape. In models which obey {\em no
  passing}~\cite{Midd92prl,MiddFish93}, the domains flipped in an avalanche
  are independent of the dynamics; eddy currents in these models won't change
  the shapes and sizes of the avalanches.}
but only for avalanches far larger and longer-lasting
than those actually seen in the experiments.

\subsubsection{Avalanche size distribution}
\label{sbsbsec:AvalHisto}

\cfig{Avalanche size distribution}
  {. The distribution of avalanche sizes in
  our model for hysteresis. Notice the logarithmic scales.%
  \protect{\footnote{We can measure a $D(S)$ value of $10^{-14}$ by 
  running billions of domains and binning over ranges $\Delta S \sim 10^5$.}}
  (i)~Although only at $R_c\approx2.16$ do we get a pure power law
  (dashed line, $D(S)\propto S^{-\bar{\tau}}$), we have large avalanches
  with hundreds of domains even a factor of two away from the
  critical point, and six orders of magnitude of scaling at 5\% above $R_c$.
  (ii)~The curves have the wrong slope except very close to the
  critical point. Be warned that a power law over two decades (although often
  publishable~\cite{MalcLidaBihaAvni97}) may not yield a reliable exponent.
  (iii)~The scaling curves (thin
  lines) work well even far from $R_c$. Inset: We plot
  $D(S) / S^{-\bar{\tau}}$ versus $S^\sigma (R-R_c)/R$ to extract
  the universal scaling curve ${\cal D}(X)$ (eqn~\ref{eq:curlyD}).
  (We use the exponent relation $\bar \tau = \tau+\sigma\beta\delta$
  without deriving it; see footnote~\ref{note:taubar} on
  page~\pageref{note:taubar}.)
  Varying the critical exponents and $R_c$ to get a good collapse allows
  us to measure the exponents far from $R_c$, where power-law fits are
  still unreliable.
  }
  {AvalHisto}{0.9\hsize}{AvalHisto.eps}

Finally, let us conclude by analyzing the avalanche size 
distribution in our RFIM for hysteresis. The
avalanche size distribution not only connects directly to many of the
experiments (Fig.~\ref{fig:DuriZapp00Sizes}), it also involves a non-obvious
example of a {\em scaling variable}).
Fig.~\ref{fig:AvalHisto} shows the distribution of
avalanche sizes $D_{\mathrm{int}}(S,R)$ for different disorders $R>R_c$,
integrated over the entire hysteresis loop
(Fig.~\ref{fig:HysteresisLoopTinyJump}).
We observe a power-law distribution near $R=R_c\sim 2.16$,
that is cut off at smaller and smaller avalanches as we move away
from $R_c$.%
  \footnote{This cutoff is the one described in section~\ref{sbsec:PowerLaws},
  scaling as $(R-R_c)^{-\sigma}$.}

Let us derive the scaling form for $D_{\mathrm{int}}(S,R)$. By using
scale invariance, we will be able to write this function of two variables
as a power of one of the variables times a universal, one-variable
function of a combined scaling variable.
From our treatment at $R_c$ (eqns~\ref{eq:DRescaleAtRc}) we know that
\begin{equation}
\begin{aligned}
S' &= S\big/\left(1+c\epsilon\right),\\
D' &= D \left(1+a\epsilon\right).
\end{aligned}
\end{equation}
A system at $R=R_c+r$ after coarse-graining will have all of its avalanches
reduced in size, and hence will appear similar to a system
further from the critical 
disorder (where the cutoff in the avalanche
size distribution is smaller, Fig.~\ref{fig:HysteresisTransition}),
say at $R=R_c+E r = R_c + (1+e \epsilon) r$. Hence
\begin{equation}
\begin{aligned}
D(S', R_c + E r) &= D'(S', R_c+r) = A D(C S', R_c + r), \\
D(S', R_c+ (1+e\epsilon) r) &=
\left(1+a\epsilon\right)
                D\left(\left(1+c\epsilon\right) S',
                                R_c + r\right).
\label{eq:DScalingEpsilon}
\end{aligned}
\end{equation}

To facilitate deriving the scaling form for multiparameter functions, it is
helpful to change coordinates so that all but one variable
remains unchanged under coarse-graining (the scaling variables). In
the average pulse shape of section~\ref{sbsbsec:PulseShape}, the time
$t$ and the duration $T$ change in the same way under coarse-graining,
so the ratio was a scaling variable. For the avalanche size 
distribution, consider the combination
$X = S^{e/c} r$. After coarse-graining $S'=S/C$ and shifting
to the higher disorder $r'=Er$ this combination is unchanged:
\begin{align}
X' = S'^{e/c} r' &= (S/C)^{e/c} (Er) =
\left(S/\left(1+c\epsilon\right) \right)^{e/c}
                        \left((1+e\epsilon) r\right) \nonumber \\
        &= S^{e/c} r
          \left(\frac{1+e\epsilon}{(1+c\epsilon)^{e/c}}\right)
             = S^{e/c} r + O(\epsilon^2) = X + O(\epsilon^2).
\end{align}
Let $\bar{D}(S,S^{e/c}R)=D(S,R)$ be the size distribution in terms of the 
variables $S$ and $X$.
Then $\bar{D}$ coarse-grains much like a function of one variable, since $X$
stays fixed. Equation~\ref{eq:DScalingEpsilon} now becomes
\begin{equation}
\label{eq:BarDScaling}
\bar{D}(S',X') = \bar{D}(S',X) =
 \left(1+a\epsilon\right)
        \bar{D}\left(\left(1+c\epsilon\right)S',X\right),
\end{equation}
so
\begin{equation}
a \bar{D} = -c S'\pp{\bar{D}}{S'}
\end{equation}
and hence
\begin{equation}
\bar{D}(S,X) = S^{-a/c} = S^{-\bar{\tau}} {\cal D}(X)
\end{equation}
with the universal {\em scaling function} ${\cal D}(X)$. 
Rewriting things in
terms of the original variables and the traditional Greek names for the
scaling exponents ($c=1/\sigma \nu$, $a = \bar{\tau}/\sigma \nu$, and
$e = 1/\nu$), we find the scaling form for the avalanche
size distribution:
\begin{equation}
\label{eq:curlyD}
D(S,R) \propto S^{-\bar{\tau}} {\cal D}(S^\sigma (R-R_c)).
\end{equation}
We can use a {\em scaling collapse} of the experimental or numerical data to
extract this universal function, by plotting $D/S^{-\bar{\tau}}$
against $X=S^\sigma (R-R_c)$; the inset of Fig.~\ref{fig:AvalHisto}
shows this scaling collapse.

In broad terms, most properties that involve large scales of length
and time at a critical point will have universal scaling forms; any
$N$ variable function will be writable in terms of a power law
times a universal function of $N-1$ variables, 
$F(x,y,z)\sim z^{-\alpha} {\cal F}(x/z^\beta, y/z^\gamma)$. The deep
significance of the renormalization-group predictions are only
feebly illustrated by the power-laws most commonly studied. The
universal scaling functions, and other morphological self-similar features
are the best ways to measure the critical exponents, the sharpest tests for
the correctness of theoretical models, and provide the richest and most
complete description of the complex behavior observed in these systems.

%
\bibliographystyle{unsrt}
\bibliography{LesHouchesSethna06}

\end{document}

%% file: JPSCommands.tex
\newcommand{\forcePageBreak}{\linebreak\clearpage\noindent}

\newcommand{\forcePageBreakNoSkip}{\linebreak\vspace*{-11pt}\clearpage\noindent}

\newcommand{\Index}[1]{}
\newcommand{\urlw}[2]{\url{#1#2}}

\newcommand{\pp}[2]{{\frac{\partial #1}{\partial #2}}}

\renewcommand{\d}[1]{{\mathrm{d}}#1}

\newcommand{\cfig}[6][t]{%
\begin{figure}
   \begin{center}
    \epsfsize=#5
    \epsffile{Fig/#6}
   \end{center}
\caption[#2]{{\bf #2}#3}
\label{fig:#4}
\end{figure}
}